\documentclass[conference,10pt]{IEEEtran}
\usepackage{cite}
\usepackage[left=0.75in,right=0.73in,top=1in,bottom=1in]{geometry}
\usepackage{amsmath,amssymb,amsfonts}
\usepackage{algorithmic}
\usepackage{graphicx}
\usepackage{textcomp}
\usepackage{xcolor}
\usepackage{cite}
\def\BibTeX{{\rm B\kern-.05em{\sc i\kern-.025em b}\kern-.08em
		T\kern-.1667em\lower.7ex\hbox{E}\kern-.125emX}}

\usepackage{nicefrac}
\usepackage{url}
\def \({\left(}
\def \){\right)}
\def \[{\left[}
\def \]{\right]}
\newcommand{\nn}{\nonumber \\}

\newcommand{\sign}{\text{ sign}}

\newcommand{\be}{\begin{equation}}
\newcommand{\ee}{\end{equation}}
\newcommand{\beqa}{\begin{eqnarray}}
\newcommand{\eeqa}{\end{eqnarray}}

\newcommand{\bea}{\begin{align}}
\newcommand{\eea}{\end{align}}

\DeclareMathAlphabet{\varmathbb}{U}{bbold}{m}{n}

\newcommand{\EE}{\mathbb{E}}

\newenvironment{talign}
 {\let\displaystyle\textstyle\align}
 {\endalign}
\newenvironment{talign*}
 {\let\displaystyle\textstyle\csname align*\endcsname}
 {\endalign}

\newtheorem{theorem}{\textbf{Theorem}}
\newtheorem{lemma}[theorem]{\textbf{Lemma}}

\newtheorem{proposition}[theorem]{\textbf{Proposition}}

\newtheorem{hypothesis}{\textbf{Hypothesis}}

\makeatletter
\def\blfootnote{\xdef\@thefnmark{}\@footnotetext}
\makeatother

\begin{document}
\title{\vspace{0.58cm}\large \textbf{CONCENTRATION OF THE MATRIX-VALUED MINIMUM MEAN-SQUARE ERROR \\IN OPTIMAL BAYESIAN INFERENCE}}
\vspace{-10mm}
\author{\IEEEauthorblockN{
		\it{Jean Barbier}\vspace{3mm}
	\IEEEauthorblockA{The Abdus Salam International Center for Theoretical Physics, Trieste, Italy.}}}
\maketitle

% \blfootnote{
% Funding from the ``Chaire de recherche sur les mod\`eles et sciences des donn\'ees'', Fondation CFM pour la Recherche-ENS is acknowledged.
% }

\begin{abstract}
We consider Bayesian inference of signals with vector-valued entries. Extending concentration techniques from the mathematical physics of spin glasses, we show that the matrix-valued minimum mean-square error concentrates when the size of the problem increases. Such results are often crucial for proving single-letter formulas for the mutual information when they exist. Our proof is valid in the \emph{optimal Bayesian inference} setting, meaning that it relies on the assumption that the model and all its hyper-parameters are known. Examples of inference and learning problems covered by our results are spiked matrix and tensor models, the committee machine neural network with few hidden neurons in the teacher-student scenario, or multi-layers generalized linear models. 
\end{abstract}
\vspace{0.2cm}

\section{\textbf{INTRODUCTION}}
This decade is witnessing a burst of mathematical studies related to inference and learning problems. One reason is that an important arsenal of methods, developed in particular in the context of spin glass physics, has found a new rich playground where it can be applied with success \cite{GhirlandaGuerra:1998,guerra2002thermodynamic,talagrand2003spin,panchenko2013sherrington}. In particular important progress has been made recently in the context of high-dimensional Bayesian inference and learning. Examples of problems in this class include spiked matrix and tensor models \cite{johnstone2001distribution,korada2009exact,deshpande2015asymptotic,krzakala2016mutual,XXT,barbier2018rank,MiolaneXX,MiolaneTensor,2017arXiv170910368B,BarbierM17a,2018arXiv180101593E,mourrat2018hamilton,Luneau_matFacto,mourrat2019hamilton}, random linear and generalized estimation \cite{KoradaMacris_CDMA,barbier_allerton_RLE,barbier_ieee_replicaCS,reeves2016replica,barbier2017phase,RLEStructuredMatrices}, models of neural networks in the teacher-student scenario \cite{barbier2017phase,SM_Arxiv_Aubin2018,Gabrie:NIPS2018}, or sparse graphical models such as error-correcting codes and block models \cite{coja2017information,abbe2018community,BarCM:2018}. 

All these results are based in some way or another on the control of the fluctuations of the {\it overlap}, which is related up to a constant to the minimum mean-square error. Optimal Bayesian inference --optimal meaning that the true posterior is known-- is an ubiquitous setting in the sense that the overlap can be shown to concentrate, and this in the whole regime of parameters (amplitude of the noise, number of observations/data points divided by the number of parameters to infer etc). 
When the overlap concentrates (and the problem is ``random enough'') one expects single-letter variational formulas for the asymptotic mutual information.

In many statistical models the overlap is a scalar. In the context of optimal Bayesian inference it is now quite standard to show that the scalar overlap is self-averaging, see, e.g., \cite{2019arXiv190106516B,BarbierM17a}. The techniques to do so are coming from communications starting with \cite{macris2007griffith,MacrisKudekar2009,KoradaMacris_CDMA} (and then generalized in \cite{andrea2008estimating,coja2017information}), and are extensions of methods used in the analysis of spin glasses \cite{GhirlandaGuerra:1998,aizenman1998stability,contucci2005spin,talagrand2003spin,panchenko2010ghirlanda}. In this paper we consider instead Bayesian inference of signals made of \emph{vectorial} components in which case the overlap is a matrix. The concentration techniques developed for scalar overlaps do not apply directly, and need to be extended using new non-trivial ideas. Examples of inference problems where matrix overlaps appear are the factorization of matrices and tensors of rank greater than one \cite{MiolaneXX}, or the committee machine neural network \cite{schwarze1992generalization,monasson1995weight,engel2001statistical,SM_Arxiv_Aubin2018}. They also appear in the context of spin glasses \cite{panchenko2018potts,panchenko2018free,agliari2018non}.
\section{\textbf{OPTIMAL INFERENCE OF TALL MATRICES}}\label{sec2}
\subsection{\textbf{General setting}}
All quantities in this paper are real. Consider a model where a ``tall'' matrix-signal $X=(X_{ik})\in [-S,S]^{n\times K}$ made of $n$ components, that are each a $K$-dimensional vector where $K\ll n$ is independent of $n$, is generated probabilistically. Its probability prior distribution $P_0$ may depend on a generic hyper-parameter $\theta_0\in \Theta_0$, i.e., $X\sim P_0(\,\cdot\,|\theta_0)$. Data (also called observations) $\widetilde Y$ are then generated conditionally on the unknown $X$ and an hyper-parameter $\theta_{\rm out}\in\Theta_{\rm out}$. Namely, the data $\widetilde{\cal Y}\ni \widetilde Y \sim P_{\rm out}(\,\cdot \,| X,\theta_{\rm out})$, with $\widetilde{\cal Y}$ a generic set: the data and hyper-parameters can be vectors, tensors etc, so that our model is very general. We assume that $\theta_0$ and $\theta_{\rm out}$ are also random, with distributions $P_{\theta_0}, P_{\theta_{\rm out}}$. 

The task is to infer the signal $X$ given the data $\widetilde Y$. We moreover assume that the hyper-parameters $\theta\equiv(\theta_0, \theta_{\rm out})$, the kernel $P_{\rm out}$ and the prior $P_0$ are known to the statistician, and call this setting {\it optimal Bayesian inference}.

The information-theoretical optimal way of infering the signal follows from its posterior distribution. Using Bayes' formula the posterior for the base inference model reads
\begin{align}
P(X=x|\widetilde Y,\theta)=\frac{P_0(x|\theta_0) P_{\rm out}(\widetilde Y|x,\theta_{\rm out})}{\int dP_0(x|\theta_0) P_{\rm out}(\widetilde Y|x,\theta_{\rm out})}\,.
% =\! \frac{P_0(x|\theta_0)}{{\cal Z}_{0,n}(\widetilde Y,\theta)} e^{-{\cal H}_0(x;\widetilde Y,\theta_{\rm out})}.
 \label{post}
\end{align}
%
% Employing the language of statistical mechanics we call 
% ${\cal H}_0(x;\widetilde Y,\theta_{\rm out})\equiv -\ln P_{\rm out}(\widetilde Y|x,\theta_{\rm out})$ the base {\it Hamiltonian}, while the posterior normalization ${\cal Z}_{0,n}(\widetilde Y,\theta)$ is the {\it partition function} of the base inference model. 
%
The averaged {\it free energy} (i.e., the Shannon entropy density of the data given the hyper-parameters) equals  
\begin{talign*}
\EE \,F_{0,n}\equiv \frac1nH(\widetilde Y|\theta)=-\frac1n \EE\ln  \int dP_0(x|\theta_0) P_{\rm out}(\widetilde Y|x,\theta_{\rm out})\,.
\end{talign*}
The average $\EE=\EE_{\theta}\EE_{X|\theta_0}\EE_{\widetilde Y|X,\theta_{\rm out}}$ is over $(\theta,X,\widetilde Y)$, jointly called the {\it quenched variables} as they are fixed by the realization of the problem, in contrast with the dynamical variable $x$ which fluctuates according to the posterior. 
% Obviously the is simply related to the mutual information density between the observations and the signal:
% %
% \begin{talign*}
% 	\frac1n I(X;\widetilde Y|\theta)= f_{0,n}-\frac1n H(\widetilde Y|X,\theta)\,.
% \end{talign*}
%
We call model \eqref{post} the ``base model'' in contrast with the perturbed model presented in section~\ref{sec:Gauss_channel}, a slightly modified version of the base model where additional side-information is given, and for which concentration results can be proved without altering the $n\to +\infty$ limit of the averaged free energy, see Lemma~\ref{lemma:same_f}.

The central object of interest is the $K\times K$ {\it overlap matrix} (or simply overlap) $Q=(Q_{kk'})$ defined as
\begin{talign*}
Q_{kk'}\equiv  \frac{1}{n} (X^\intercal x)_{kk'}= \frac{1}{n}	\sum_{i=1}^nX_{ik}x_{ik'}\,.
\end{talign*}
Here $x$ is a sample from the posterior and $X$ is the signal. The overlap contains a lot of information. Using that the estimator minimizing the mean-square error is the posterior mean $\langle x \rangle_0\equiv \EE[X|\widetilde Y,\theta]$ (denoting $\langle - \rangle_0$ the expectation w.r.t. the posterior \eqref{post} of the base model), the $K\times K$ matrix-valued minimum mean-square error (MMSE) is
\begin{talign*}
\frac1n \EE\big[(X-\langle x\rangle_0)^\intercal (X-\langle x\rangle_0)\big]=\EE[X_1X_1^\intercal]-\EE\langle Q\rangle_0
\end{talign*}
where $X_i^\intercal \in\mathbb{R}^K$ is a row of $X$ (all vectors are columns, including rows of matrices considered alone, transposed vectors are rows). The scalar MMSE is then simply 
\begin{talign*}
\frac1n\EE\|X-\langle x\rangle_0\|^2_{\rm F}= \EE\|X_1\|^2 - {\rm Tr}\,\EE\langle Q\rangle_0
\end{talign*}
($\rm Tr$ is the trace). Another metric for problems where, e.g., the sign of the signal is lost due to symmetries is
\begin{talign*}
\frac{1}{n^2}\EE\|X X^\intercal\!-\! \langle x x^\intercal\rangle_0\|^2_{\rm F} = \EE\big[(X_1^\intercal X_2)^2\big] \!-\! \EE\big\langle \|Q\|^2_{\rm F}\big\rangle_0\! +\! {\cal O}(\frac1n).
\end{talign*}
\subsection{\textbf{Examples}}\label{secEx}
In the symmetric order-$p$ rank-$K$ tensor factorization problem, the data-tensor $\widetilde Y=(\widetilde Y_{i_1\ldots i_p})$ is generated as 
\begin{talign}
\widetilde Y_{i_1\ldots i_p}=n^{\frac{1-p}{2}}\,\sum_{k=1}^K X_{i_1 k}X_{i_2 k}\ldots X_{i_p k} + \widetilde Z_{i_1\ldots i_p} \label{ex:tensorFacto}
\end{talign}
for $1\le i_1\le i_2\le \ldots \le i_p\le n$. Here $\widetilde Z$ is a Gaussian noise tensor with independent and identically distributed (i.i.d.) ${\cal N}(0,1)$ entries, and the signal components are i.i.d., i.e., with a prior $P_0=p_0^{\otimes n}$ with $p_0$ supported on $[-S,S]^K$. The case $p=2$ is the spiked Wigner model and is one of the simplest probabilistic model for principal component analysis \cite{johnstone2001distribution}. In both the analysis of \cite{Luneau_matFacto,mourrat2019hamilton} for this problem the matrix overlap concentration is a key result. 

Another model is the following generalized linear model (GLM) (recall $X_i\in \mathbb{R}^K$):
\begin{talign}
\widetilde Y_\mu \sim p_{\rm out}\big(\cdot \big| \sum_{i=1}^n \theta_{\mu i}X_{i}\big), \qquad 1\le\mu\le m=\Theta(n). \label{ex:GLM}
\end{talign}
Given $\mathbb{R}^{m\times n}\ni\theta_{\rm out}=(\theta_\mu)_{\mu=1}^m$ and $X$, the $m$ data points are i.i.d., thus the notation $p_{\rm out}$ instead of $P_{\rm out}(\,\cdot\,|\theta_{\rm out} X)=\otimes_{\mu=1}^m p_{\rm out}(\,\cdot\,|X^\intercal \theta_\mu)$. We also assume that the prior $P_0=p_0^{\otimes n}$. A particular simple deterministic case is 
\begin{talign}
\widetilde Y_\mu ={\sign}\sum_{k=1}^K {\sign}\sum_{i=1}^n \theta_{\mu i}X_{ik}\,, \quad 1\le\mu\le m\,. \label{committee}
\end{talign}
This model is a version of the committee machine \cite{barbier2017phase,SM_Arxiv_Aubin2018}. Here $(X_{ik})_{i=1}^n$ can be interpreted as the weights of the $k$-th hidden neuron, and $(\theta_\mu)$ are $n$-dimensional data points used to generate the labels $(\widetilde Y_\mu)$. The teacher-student scenario in which our results apply is: the teacher network \eqref{committee} (or \eqref{ex:GLM} in general) generates $\widetilde Y$ from the data $\theta_{\rm out}$. The pairs $(\widetilde Y_{\mu}, \theta_\mu)$ are then used in order to train (i.e., learn the weights of) a student network with the same architecture.

A richer example is a multi-layer version of the GLM:
\begin{align}\label{ex:multiLayer}
\begin{cases}
X^{(L)}_{i_L} \sim p_{\rm out}^{(L)}\big(\cdot \big| \sum_{j=1}^{n_{L-1}} \theta_{i_L j}^{(L)}X_{j}^{(L-1)}\big),\\
X_{i_{L-1}}^{(L-1)}\sim  p_{\rm out}^{(L-1)}\big(\cdot \big| \sum_{j=1}^{n_{L-2}} \theta_{i_{L-1} j}^{(L-1)}X_{j}^{(L-2)}\big),\\
\qquad\qquad\qquad\qquad\vdots\\
X_{i_1}^{(1)}\sim  p_{\rm out}^{(1)}\big(\cdot \big| \sum_{j=1}^{n_0} \theta_{i_1 j}^{(1)}X_{j}^{(0)}\big),
\end{cases}
\end{align}
where each index $i_\ell$ runs from $1$ to $n_{\ell}=\Theta(n_0)$, for $\ell=1,\ldots,L$. The input $X^{(0)}\sim P_{0}$ is factorized as $P_0=p_0^{\otimes n_0}$. In this model $(X^{(\ell)})_{\ell=1}^{L-1}$ represent intermediate hidden variables, the visible variable $X^{(L)}=\widetilde Y$ is the data, and $\theta_{\rm out}=(\theta^{(\ell)})$ with $\theta^{(\ell)}$ representing the weight matrix at the $\ell$-th layer. Note that in the single layer case \eqref{ex:GLM}, $\theta_{\rm out}$ was interpreted as data points and $X$ as the weight vector to learn. This model has been studied by various authors when $K=1$ and when the components $(X_{j}^{(\ell)})$ are scalars \cite{manoel_multi-layer_2017,reeves_additivity_2017,fletcher_inference_2017,Gabrie:NIPS2018,DBLP:journals/corr/abs-1903-01293}. But one can define generalizations where these are multi-dimensional, in which case overlap matrices naturally appear.

A last example could be another combinaison of statistical models such as a spiked Wigner model where the hidden low-rank representation $X$ of the data $\widetilde Y$ has a complex generative prior. For example $X$ could be generated from a GLM over a more primitive signal $X^{(0)}\sim P_0=p_0^{\otimes n_0}$ (here $n=\Theta(n_0)$):
\begin{align*}
\begin{cases}
\widetilde Y_{ij}=n^{-1/2}\,\sum_{k=1}^K X_{ik}X_{jk}+Z_{ij}\,,  &1\le i\le j\le n\,,\\
X_i \sim p_{\rm out}\big(\cdot \big| \sum_{j=1}^{n_0} \theta_{i j}X_{j}^{(0)}\big)\,, &1\le i\le n\,.	
\end{cases}
\end{align*}
This set-up has recently attracted attention \cite{aubin2019spiked} for studying models of complex structured data with generative priors.

\section{\textbf{THE PERTURBED MODEL, AND RESULTS}}\label{sec:Gauss_channel}
\subsection{\textbf{The vectorial Gaussian channel perturbation}}
In order to ``force'' overlap concentration we need, in addition to the data $\widetilde Y$, infinitesimal side-information $Y\in\mathbb{R}^{n\times K}$ about $X$ coming from a vectorial Gaussian channel:
\begin{align}
Y = X \lambda_n^{1/2} + Z, \  \text{i.e.,}  \ Y_i = \lambda_n^{1/2}\, X_i +  Z_i, \  1\le i\le n.\label{pert_channel}
\end{align}
The i.i.d. Gaussian noise $(Z_i)\sim {\cal N}(0,I_K)^{\otimes n}$. The signal-to-noise (SNR) {\it matrix} controlling the signal strength $\lambda_n\equiv s_n \tilde \lambda$ with a sequence $(s_n)$ that tends to $0_+$, and $\tilde\lambda$ belongs to 
\begin{talign*}
{\cal D}_{K} \equiv \,&\big\{\tilde\lambda\in\mathbb{R}^{K\times K} : \tilde\lambda_{kk'}=\tilde\lambda_{k'k}\in(1,2) \ \forall \ k\neq k', \\
&\qquad\qquad\qquad\qquad\qquad\tilde\lambda_{kk}\in(2K,2K+1) \ \forall\ k  \big\}	
\end{talign*}
(other sets could be used but this one is convenient for the proof). We also denote ${\cal D}_{n,K}\equiv s_n {\cal D}_K$ so that $\lambda_n\in {\cal D}_{n,K}$. Matrices belonging to ${\cal D}_{n,K}$ are symmetric strictly diagonally dominant with positive entries and thus ${\cal D}_{n,K}\subset {\cal S}_{K}^+$, where ${\cal S}_{K}^+$ is the set of symmetric positive definite matrices of dimension $K \times K$, see \cite{horn1990matrix}. As $\lambda_n\in {\cal D}_{n,K}$ it has a unique square root matrix that we denote $\lambda_n^{1/2}=\sqrt{s_n}\,\tilde\lambda^{1/2}$.

The \emph{perturbed inference model} is then
\begin{align}\label{2channels}
	\begin{cases}
	\widetilde Y  \!\!\!\!\! &\sim P_{\rm out}(\,\cdot\, |X,\theta_{\rm out}),\\
	Y \!\!\!\!\!&= X \lambda_n^{1/2} + Z.
	\end{cases}
\end{align}
It is called ``perturbed model'' because the base model has been slightly modified by adding new data points coming from \eqref{pert_channel} that are ``weak'' (as $s_n\to 0_+$).
%  Let 
% %
% \begin{talign*}
% {\cal H}_{\lambda_n}(x;Y(X,Z))\!\equiv\!\sum_{i=1}^{n}\! \big( \frac {1}{2}x_i^\intercal\lambda_n x_i - x_{i}^\intercal \lambda_n X_{i}-x_{i}^\intercal \lambda_n^{1/2} \, Z_{i}\big).
% \end{talign*} 
% %
% The total Hamiltonian is therefore the sum of the base Hamiltonian and the perturbation one. 
The posterior $P(X=x|\widetilde Y,Y,\theta,\lambda_n)$ of the perturbed model reads
\begin{align}
&\frac{P_0(x|\theta_0) P_{\rm out}(\widetilde Y|x,\theta_{\rm out})  e^{-{\cal H}_{\lambda_n}(x,Y)}}{\int dP_0(x|\theta_0) P_{\rm out}(\widetilde Y|x,\theta_{\rm out}) e^{-{\cal H}_{\lambda_n}(x,Y)}}. \label{post_pert}
\end{align}
where ${\cal H}_{\lambda_n}(x,Y)\equiv \frac12\|x\lambda_n^{1/2}\|_{\rm F}^2-{\rm Tr}(Y^\intercal x\lambda_n^{1/2})$. We define the bracket $\langle -\rangle$ as the expectation w.r.t. the posterior of the perturbed model: $\langle g\rangle \equiv \int dP(X=x|\widetilde Y,Y,\theta,\lambda_n)\, g(x)$. Thus $\langle g\rangle$ depends on $(\widetilde Y,Y,\theta)$ and the SNR $\lambda_n$.

It is crucial to notice that the perturbed model \eqref{2channels} is set in the optimal Bayesian inference setting. Again, this means that in addition to the data $(\widetilde Y,Y)$ the statistician knows the data generating model, namely the kernel $P_{\rm out}$ and the additive Gaussian nature of the noise in the second channel in \eqref{2channels}, the prior $P_0$ as well as all hyper-parameters $(\theta,\lambda_n)$, and is therefore able to write the true posterior \eqref{post_pert}. 

An important object is the free energy of model \eqref{2channels}:
\begin{talign*}
F_{n}(\lambda_n)\equiv-\frac{1}{n}\ln \int dP_0(x|\theta_0) P_{\rm out}(\widetilde Y|x,\theta_{\rm out})  e^{-{\cal H}_{\lambda_n}(x,Y)}.
\end{talign*}
Concentration of the overlap requires an hypothesis:
\begin{hypothesis}[Free energy concentration]\label{hyp1}
	There exists a constant $C_F$ that may depend on everything but $n$, and s.t.
	\begin{talign}
	\EE\big[\big(F_n(\lambda_n)-\EE \, F_n(\lambda_n)\big)^2\big]\le C_F\,n^{-1}\,. \label{hyp:f_conc}
	\end{talign}
\end{hypothesis} 

The expectation $\mathbb{E}\equiv \mathbb{E}_\theta\mathbb{E}_{X|\theta_0}\mathbb{E}_{\widetilde{Y}|X,\theta_{\rm out}}\mathbb{E}_{Y|X,\lambda_n}$ is over all quenched variables but not over $\lambda_n$, which remains fixed. 

For purely generic optimal inference models without any restrictions on the distributions $(P_0,P_{\rm out},P_{\theta_0},P_{\theta_{\rm out}})$ it is generally very hard, if not wrong, to try proving \eqref{hyp:f_conc}. The model must be ``random enough'' and possess some underlying factorization structure for such hypothesis to be true (thus the factorization properties assumed in the examples of section \ref{secEx}). The most studied case in the literature is when the prior and the kernel factorize, namely $P_0=p_0^{\otimes n}$ and the data points are i.i.d. given $(X,\theta_{\rm out})$. The examples \eqref{ex:tensorFacto}--\eqref{committee} fall in this class. Under such factorization assumptions it is quite straightforward to prove \eqref{hyp:f_conc} using standard techniques (see, e.g., \cite{BarbierM17a,barbier2017phase}). But such simple factorization properties are not always there, as illustrated by the two last examples in section~\ref{secEx}. In these examples it is a perfectly valid question to wonder wether the overlap of the hidden variables do concentrate\footnote{Note that proving concentration of the overlap for a hidden variable requires a perturbation of the form \eqref{pert_channel} over the hidden variable, not over $X^{(0)}$, which in this case is just interpreted as a constitutive element of the prior of the hidden variable of interest, see \cite{Gabrie:NIPS2018} where this is done.} (this question is crucial in the analysis of \cite{Gabrie:NIPS2018}). The hidden variables have very complex structured prior (i.e., probability distribution), with highly non-trivial factorization properties, in which case proving \eqref{hyp:f_conc} requires work. See, e.g., \cite{Gabrie:NIPS2018} where this has been done for the multi-layer GLM \eqref{ex:multiLayer} with a single hidden layer ($L=2$), where this is already challenging. 

An important feature of the perturbation is that it does not change the limit of the averaged free energy; this means that in a certain sense the perturbed model is equivalent to the base one at a ``macroscopic'' level, i.e., for the global quantities. We denote in this paper $C$ a generic constant that may depend on all parameters in the problem like $K$ and $S$ but not on $n$.

\begin{lemma}[Free energy equivalence]\label{lemma:same_f} There exists a constant $C$ s.t. $|\EE\,F_{0,n} - \EE\,F_{n}(\lambda_n) |	\le Cs_n$. Thus $\EE\, F_{0,n}$ and $\EE\,F_{n}(\lambda_n)$ have same $n\to+\infty$ limit, provided it exists.
\end{lemma}
\subsection{\textbf{Main results}}
% In order to give our first result we need to introduce the overlap between two replicas $$Q^{(12)}\equiv\frac1n \sum_{i=1}^n x^{(1)}_i(x_i^{(2)})^\intercal\,,$$ where, again, replicas are i.i.d. random variables drawn accroding to the posterior measure \eqref{post_pert} of the perturbed model (and thus share the same quenched variables): $(x^{(1)},x^{(2)})\sim P(\cdot|\widetilde Y,Y,\theta,\lambda_n)^{\otimes 2}$. By a slight abuse of notation let us continue to use the same bracket notation for the expectation of functions of replicas w.r.t. to the product posterior measure: $$\big\langle g(x^{(1)},x^{(2)})\big\rangle\equiv\int dP(x^{(1)}|\widetilde Y,Y,\theta,\lambda_n)dP(x^{(2)}|\widetilde Y,Y,\theta,\lambda_n) g(x^{(1)},x^{(2)})\,.$$ 
%
Our main results are concentration theorems for the overlap in a (perturbed) model of optimal Bayesian inference. We start with the first type of fluctuations, namely the fluctuations of the overlap w.r.t. the posterior distribution, or what is called ``thermal fluctuations'' in statistical mechanics. Controlling these fluctuations does not require that the free energy concentrates (the hypothesis \eqref{hyp:f_conc} is not required). 
Denote $\EE_{\lambda} \,g \equiv {\rm Vol}({\cal D}_{n,K})^{-1}\int_{{\cal D}_{n,K}} d\lambda_n\,g(\lambda_n)$
% \begin{talign*}
% \EE_{\lambda} \,g \equiv {\rm Vol}({\cal D}_{n,K})^{-1}\int_{{\cal D}_{n,K}} d\lambda_n\,g(\lambda_n)	
% \end{talign*}
the average over the perturbation matrix $\lambda_n$, where ${\rm Vol}({\cal D}_{n,K})=s_n^{K(K+1)/2}$ ($\lambda_n$ has $K(K+1)/2$ independent entries). Then:
\begin{theorem}[Thermal fluctuations of $Q$]\label{prop:thermalBOund}
Consider an optimal Bayesian inference problem (i.e., for which the true posterior is known), with side information coming from the channel \eqref{pert_channel}; i.e., a model of the form \eqref{2channels}. Let $(s_n)$ a sequence s.t. $s_n\to 0_+$ and $s_nn\to+\infty$. There exists $C>0$ s.t.
\begin{talign*}
\EE_\lambda\EE\big\langle \|Q- \langle Q\rangle\|_{\rm F}^2 \big\rangle &\le C(s_nn)^{-1/2}.
% \EE_\lambda\EE\big\langle \|Q-\langle Q^{(12)}\rangle\|_{\rm F}^2	\big\rangle   &\le \frac{C(K,S)}{\sqrt{s_nn}} \,.\label{2terms_controlled}
\end{talign*}
\end{theorem}

The next, stronger, result takes care of the additional fluctuations due to the quenched randomness, and requires this time the free energy concentration hypothesis:
\begin{theorem}[Total fluctuations of $Q$]\label{thm:Q_con}
Consider a perturbed optimal Bayesian inference problem of the form  \eqref{2channels}. Assume Hypothesis~\ref{hyp1}. Let $(s_n)$ verify $s_n\to 0_+$ and $s_n^4n\to+\infty$. There exists $C=C(C_f,K,S)>0$ s.t.
\begin{talign*}
\EE_\lambda\EE\big\langle \|Q-\EE\langle Q\rangle\|_{\rm F}^2\big\rangle \le C(s_n^4n)^{-1/6}\,.
\end{talign*}	
\end{theorem}

We emphasize that, as Theorem~\ref{prop:thermalBOund} does not require Hypothesis~\ref{hyp1}, it is valid very generically, even for very complex models without any factorization properties for the signal's prior nor for the kernel; it is only a consequence of the perturbation and the Bayesian optimality. In such models, deriving single-letter formulas for quantities like the mutual information or the MMSE is doomed (as generally there is not). Indeed, proofs of such simple formulas always require in one way or another strong factorization properties, directly, like, e.g., in \cite{MiolaneXX,MiolaneTensor}, or indirectly through the need of the stronger concentration result Theorem~\ref{thm:Q_con} as in \cite{BarbierM17a,2017arXiv170910368B,barbier2017phase,Luneau_matFacto,mourrat2018hamilton,mourrat2019hamilton}\footnote{In these papers the concentration needs to be proven for an appropriate ``interpolating'' model.}. 

Another remark is related to the role of the perturbation (i.e., side-information). Our theorems require an external average $\EE_\lambda$ over the perturbation: this is \emph{not} an artefact of the proof. Indeed, there might be a (zero-measure) set in the hyper-parameters space $\Theta_0\times \Theta_{\rm out}$ where, in the $n\to+\infty$ limit, there are \emph{phase transitions}. Phase transitions manifest themselves in particular by a non self-averaging behavior of the overlap. But averaging over a vanishing window of $\lambda_n$, which importantly is independent of $\theta$, allows to ``smoothen'' the overlap fluctuations, effectively cancelling the dramatic effect of possible phase transitions.

% Before entering the proof let us make a very last remark. There are problems with multiple overlaps. For example one may also consider the non-symmetric version of the tensor factorization problem. In this case $p$ matrices $X^{[p]}\in \mathbb{R}^{n_p\times K}$, with $n_p=\Theta(n)$ and with a possibly matrix-dependent prior $P_0^{[p]}$, are to be reconstructed from a data-tensor of the form
% %
% \begin{align*}
% \widetilde Y_{i_1\ldots i_p}=n^{\frac{1-p}{2}}\,\sum_{k=1}^K X_{i_1 k}^{[1]}X_{i_2 k}^{[2]}\ldots X_{i_p k}^{[p]} + \widetilde Z_{i_1\ldots i_p}\,, \qquad 1\le i_1, i_2, \ldots, i_p\le n\,.
% \end{align*}
% %
% In this case there is one overlap per matrix-signal to be inferred: $$Q^{[p]} \equiv \frac1{n_p} \sum_{i=1}^{n_p} X_{i}^{[p]}(x_i^{[p]})^\intercal\,.$$ It should be clear to the reader that all the setting described in this paper can be straightforwardly extended to include this case: one has to consider one perturbation channel of the form \eqref{pert_channel} per variable to be reconstructed (i.e., per matrix in the non-symmetric tensor factorization problem), each with its own independent matrix SNR: $$Y_i^{[p]} = (\lambda_n^{[p]})^{1/2}\, X_i^{[p]} +  Z_i^{[p]}\,, \qquad 1\le i\le n_p\,.$$ Then the total Hamiltonian is the sum of the base one and the $p$ perturbation Hamiltonians, and so forth.
\section{\textbf{PROOF IDEA}}
We give few pointers to help the reader to get idea of the proof. All details can be found in \cite{barbier2019overlap}. 

Proving overlap concentration relies on the concentration of another $K\times K$ matrix ${\cal L}\equiv\frac1n\nabla_{\lambda_n} {\cal H}_{\lambda_n}(x,Y)$.

\begin{proposition}[Concentration of ${\cal L}$]\label{prop:concL}
Let $s_n\to 0_+$ and $s_nn\to+\infty$. Then there is $C>0$ s.t.
\begin{talign}
\EE_\lambda\EE\big\langle \|{\cal L}- \langle {\cal L}\rangle\|_{\rm F}^2 \big\rangle &\le C(s_nn)^{-1}\,.\label{thL}
% \EE_\lambda\EE\big\langle \|Q-\langle Q^{(12)}\rangle\|_{\rm F}^2	\big\rangle   &\le \frac{C(K,S)}{\sqrt{s_nn}} \,.\label{2terms_controlled}
\end{talign}
If $s_n^4n\to+\infty$ and Hypothesis~\ref{hyp1} is verified, 
\begin{talign}
\EE_\lambda\EE\big\langle \|{\cal L}-\EE\langle {\cal L}\rangle\|_{\rm F}^2\big\rangle \le C(s_n^4n)^{-1/3}\,.\label{disL}
\end{talign}	
\end{proposition}

The fluctuations of this matrix are easier to control than the ones of the overlap because ${\cal L}$ is related to the $\lambda_n$-gradient of the free energy, which is self-averaging by hypothesis \eqref{hyp:f_conc}. The proof is a straightforward extension to the matrix case of the one found in \cite{2019arXiv190106516B,barbier2017phase} and requires no new ideas. This general result does not depend on the fact that we consider optimal Bayesian inference; it is only a consequence of the perturbation, i.e., the side information coming from the channel \eqref{pert_channel}. What instead \emph{does} require new ideas and relies on the Bayesian optimal setting is the link between the concentration of ${\cal L}$ and the one of $Q$. The additional difficulty w.r.t. what is done in \cite{2019arXiv190106516B,barbier2017phase} for a scalar overlap (i.e., the case $K=1$) is that the matrix $Q$ is not symmetric, even if its expectation $\mathbb{E}\langle Q\rangle$ is. Symmetry in expectation is a consequence of the general identity (sometimes called ``Nishimori identity'') $\EE\big\langle g(x,X;\widetilde Y,Y)\big\rangle =\EE\big\langle g(x,x';\widetilde Y,Y)\big\rangle$, where $X$ is the signal, $x,x'$ are i.i.d. samples from the posterior \eqref{post_pert}, $\langle -\rangle$ is the expectation w.r.t. the product posterior measure, and $g$ is any bounded function. This innocent-looking key identity on which relies the whole proof follows directly from Bayes' law --thus the importance of placing ourselves in the Bayesian optimal setting--, see \cite{2019arXiv190106516B,barbier2017phase}. Applied to $\mathbb{E}\langle Q\rangle=\mathbb{E}[X^\intercal\langle  x\rangle]=\mathbb{E}\langle (x')^\intercal x\rangle=\mathbb{E}[\langle x\rangle^\intercal \langle x\rangle]$ which is indeed symmetric. 
% \subsection{\textbf{Linking the thermal fluctuations of $\cal L$ and $Q$}}

\textbf{Linking the thermal fluctuations of $\cal L$ and $Q$}:
Let us start by giving the main steps behind the proof of Theorem~\ref{prop:thermalBOund}. The key insight is the following inequality: by definition of the overlap (and for any $l,l'\in\{1,\ldots,K\}$),
\begin{talign}
&\EE_\lambda\EE\big\langle (Q_{ll'}- \langle Q_{ll'}\rangle)^2 \big\rangle=\EE_\lambda\EE\langle Q_{ll'}^2\rangle- \EE_\lambda\EE\big[\langle Q_{ll'}\rangle^2\big]\label{toUse}\\
&=\frac1{n^2}\sum_{i,j=1}^n\EE_\lambda\EE\big[ X_{il}X_{jl} (\langle x_{il'}x_{jl'}\rangle-\langle x_{il'}\rangle\langle x_{jl'}\rangle)\big]\nn
&\le C \big\{\frac1{n^2}\sum_{i,j=1}^n\EE_\lambda\EE\big[(\langle x_{il'}x_{jl'}\rangle-\langle x_{il'}\rangle\langle x_{jl'}\rangle)^2\big] \big\}^{1/2}\nonumber
\end{talign}
for some $C>0$ using Cauchy-Schwarz, and that the prior has bounded support. Combining the Nishimori identity, Gaussian integration by parts and by careful algebra using the formula $\frac{d\lambda}{d\lambda_{ll'}}=\lambda^{1/2}\frac{d\lambda^{1/2}}{d\lambda_{ll'}} + \frac{d\lambda^{1/2}}{d\lambda_{ll'}}\lambda^{1/2}$ one can show
\begin{talign*}
\frac{1}{n^2}\sum_{i,j=1}^n \EE\big[(\langle x_{il}x_{jl}\rangle -\langle x_{il}\rangle \langle x_{jl}\rangle)^2\big]&\!\le \!2\mathbb{E}\big\langle (\mathcal{L}_{ll} - \langle \mathcal{L}_{ll}\rangle)^2\big\rangle \nn
&\quad +C(s_nn)^{-1}\,.
\end{talign*}
Identity \eqref{thL} in Proposition \eqref{prop:concL} for $\cal L$ then implies that the $x_{il}$'s asymptotically ``decouple''. When this is plugged in \eqref{toUse}, this decoupling property translates into Theorem~\ref{prop:thermalBOund}.
%
% \subsection{\textbf{Total fluctuations of $Q$}}
%

\textbf{Total fluctuations of $Q$}:
We now consider Theorem~\ref{thm:Q_con}, which requires to obtain Theorem~\ref{prop:thermalBOund} first. The proof ressembles the derivation of the Ghirlanda-Guerra identities in the context of spin glasses \cite{GhirlandaGuerra:1998}. The main identity is
\begin{talign}
&\EE_\lambda{\rm Tr}\,\EE \big\langle Q({\cal L} -\EE\langle {\cal L}\rangle) \big\rangle	\!=\! -\sum_{l\neq l'}\EE_\lambda\EE\big\langle (Q_{ll'}-\EE\langle Q_{ll'}\rangle)^2\big\rangle\nn
&- \frac12\sum_{l}\EE_\lambda\EE\big\langle (Q_{ll}-\EE\langle Q_{ll}\rangle)^2\big\rangle \! +\!\EE_\lambda{\rm Tr}\,\EE\big\langle Q(\langle Q^{(12)}\rangle -Q)\big\rangle\nn
&+  \frac12\sum_{l}\EE_\lambda\EE\big\langle Q_{ll}(Q_{ll}- \langle Q_{ll}^{(12)}\rangle)\big\rangle \pm C(s_nn)^{-1/4}. \label{mainId} 
\end{talign}
Here $Q^{(12)}\equiv x^\intercal x'$ is the overlap between two i.i.d. samples from the (perturbed) posterior \eqref{post_pert}, so that $\langle Q^{(12)}\rangle = \langle x\rangle^\intercal \langle x\rangle$ is symmetric. The relation \eqref{mainId} is shown by the Nishimori identity and Gaussian integration by parts, which in particular allows to prove $\EE\langle {\cal L}\rangle = \frac12 {\rm diag}(\EE\langle Q\rangle)-\EE\langle Q\rangle$, as well as the use of Theorem~\ref{prop:thermalBOund}. Again, in the derivation one has to be careful as the matrices that appear are not symmetric, which complicates the task. 

Now, by \eqref{disL} in Proposition \eqref{prop:concL} and the Cauchy-Schwarz inequality we have that the left hand side of \eqref{mainId} verifies 
\begin{talign*}
\big|\EE_\lambda{\rm Tr}\,\EE \big\langle Q({\cal L} -\EE\langle {\cal L}\rangle) \big\rangle\big|&\le C\big\{\EE_\lambda\EE \big\langle \|{\cal L} -\EE\langle {\cal L}\rangle\|_{\rm F}^2 \big\rangle\big\}^{1/2}\nn
&\le C(s_n^4n)^{-1/6}	.
\end{talign*}
Therefore, because of the alternating signs on the right hand side of \eqref{mainId}, showing that $\EE_\lambda\EE\big\langle \|Q-\EE\langle Q\rangle\|_{\rm F}^2\big\rangle$ is small requires to prove that the third and forth terms are small. These can be thought of as a ``measure of asymmetry'' of the overlap matrix $Q$. The last crucial step is therefore showing (this again relies on the Nishimori identity)
\begin{talign*}
\EE_\lambda\EE\big\langle \|Q-\langle Q^{(12)}\rangle\|_{\rm F}^2	\big\rangle   &\le C(s_nn)^{-1/2}.
\end{talign*}

\section*{\textbf{ACKNOWLEDGMENTS}}
Funding from Fondation CFM pour la Recherche-ENS is acknowledged.
I would like to thank Nicolas Macris, Dmitry Panchenko, Antoine Maillard, Florent Krzakala, L\'eo Miolane and Cl\'ement Luneau for discussions.
\bibliographystyle{unsrt_abbvr}
\bibliography{camsap19}

\begin{thebibliography}{10}

\bibitem{GhirlandaGuerra:1998}
S.~Ghirlanda and F.~Guerra.
\newblock General properties of overlap probability distributions in disordered
  spin systems. towards parisi ultrametricity.
\newblock {\em Journal of Physics A: Mathematical and General}, 31(46):9149,
  1998.

\bibitem{guerra2002thermodynamic}
F.~Guerra and F.~L. Toninelli.
\newblock The thermodynamic limit in mean field spin glass models.
\newblock {\em Communications in Mathematical Physics}, 230(1):71--79, 2002.

\bibitem{talagrand2003spin}
M.~Talagrand.
\newblock {\em Spin glasses: a challenge for mathematicians: cavity and mean
  field models}, volume~46.
\newblock Springer, 2003.

\bibitem{panchenko2013sherrington}
D.~Panchenko.
\newblock {\em The Sherrington-Kirkpatrick model}.
\newblock Springer Science \& Business Media, 2013.

\bibitem{johnstone2001distribution}
I.~Johnstone.
\newblock On the distribution of the largest eigenvalue in principal components
  analysis.
\newblock {\em The Annals of statistics}, 29(2):295--327, 2001.

\bibitem{korada2009exact}
S.~B. Korada and N.~Macris.
\newblock Exact solution of the gauge symmetric p-spin glass model on a
  complete graph.
\newblock {\em Journal of Statistical Physics}, 136(2):205--230, 2009.

\bibitem{deshpande2015asymptotic}
Y.~Deshpande, E.~Abbe, and A.~Montanari.
\newblock Asymptotic mutual information for the balanced binary stochastic
  block model.
\newblock {\em Information and Inference: A Journal of the IMA}, 6(2):125--170,
  2016.

\bibitem{krzakala2016mutual}
F.~Krzakala, J.~Xu, and L.~Zdeborov{\'a}.
\newblock Mutual information in rank-one matrix estimation.
\newblock In {\em 2016 IEEE Information Theory Workshop (ITW)}, pages 71--75,
  Sept 2016.

\bibitem{XXT}
J.~Barbier, M.~Dia, N.~Macris, F.~Krzakala, T.~Lesieur, and L.~Zdeborov\'a.
\newblock Mutual information for symmetric rank-one matrix estimation: A proof
  of the replica formula.
\newblock In {\em Advances in Neural Information Processing Systems 29}, page
  424–432. 2016.

\bibitem{barbier2018rank}
J.~Barbier, M.~Dia, N.~Macris, F.~Krzakala, and L.~Zdeborov{\'a}.
\newblock Rank-one matrix estimation: analysis of algorithmic and information
  theoretic limits by the spatial coupling method.
\newblock {\em arXiv:1812.02537}, 2018.

\bibitem{MiolaneXX}
M.~Lelarge and L.~Miolane.
\newblock Fundamental limits of symmetric low-rank matrix estimation.
\newblock {\em Probability Theory and Related Fields}, 173(3-4):859--929, 2019.

\bibitem{MiolaneTensor}
T.~Lesieur, L.~Miolane, M.~Lelarge, F.~Krzakala, and L.~Zdeborov{\'a}.
\newblock Statistical and computational phase transitions in spiked tensor
  estimation.
\newblock In {\em 2017 IEEE International Symposium on Information Theory
  (ISIT)}, pages 511--515. IEEE, 2017.

\bibitem{2017arXiv170910368B}
J.~{Barbier}, N.~{Macris}, and L.~{Miolane}.
\newblock {The Layered Structure of Tensor Estimation and its Mutual
  Information}.
\newblock In {\em 55th Annual Allerton Conference on Communication, Control,
  and Computing}, 2017.

\bibitem{BarbierM17a}
J.~Barbier and N.~Macris.
\newblock The adaptive interpolation method: a simple scheme to prove replica
  formulas in bayesian inference.
\newblock {\em Probability Theory and Related Fields}, Oct 2018.

\bibitem{2018arXiv180101593E}
A.~{El Alaoui} and F.~{Krzakala}.
\newblock {Estimation in the Spiked Wigner Model: A Short Proof of the Replica
  Formula}.
\newblock In {\em IEEE International Symposium on Information Theory (ISIT)},
  2017.

\bibitem{mourrat2018hamilton}
J.-C. Mourrat.
\newblock Hamilton-jacobi equations for mean-field disordered systems.
\newblock {\em preprint arXiv:1811.01432}, 2018.

\bibitem{Luneau_matFacto}
J.~Barbier, C.~Luneau, and N.~Macris.
\newblock Mutual information for low-rank even-order symmetric tensor
  factorization.
\newblock In {\em 2019 IEEE Information Theory Workshop}.

\bibitem{mourrat2019hamilton}
J.-C. Mourrat.
\newblock Hamilton-jacobi equations for finite-rank matrix inference.
\newblock {\em preprint arXiv:1904.05294}, 2019.

\bibitem{KoradaMacris_CDMA}
S.~B. Korada and N.~Macris.
\newblock Tight bounds on the capacity of binary input random {CDMA} systems.
\newblock {\em IEEE Trans. on Information Theory}, 56(11):5590--5613, Nov 2010.

\bibitem{barbier_allerton_RLE}
J.~Barbier, M.~Dia, N.~Macris, and F.~Krzakala.
\newblock {The Mutual Information in Random Linear Estimation}.
\newblock In {\em in the 54th Annual Allerton Conference on Communication,
  Control, and Computing}, 2016.

\bibitem{barbier_ieee_replicaCS}
J.~Barbier, N.~Macris, M.~Dia, and F.~Krzakala.
\newblock Mutual information and optimality of approximate message-passing in
  random linear estimation.
\newblock {\em preprint arXiv:1701.05823}, 2017.

\bibitem{reeves2016replica}
G.~Reeves and H.~D. Pfister.
\newblock The replica-symmetric prediction for compressed sensing with gaussian
  matrices is exact.
\newblock In {\em IEEE International Symposium on Information Theory (ISIT)},
  pages 665--669, 2016.

\bibitem{barbier2017phase}
J.~Barbier, F.~Krzakala, N.~Macris, L.~Miolane, and L.~Zdeborov{\'a}.
\newblock Optimal errors and phase transitions in high-dimensional generalized
  linear models.
\newblock {\em Proceedings of the National Academy of Sciences},
  116(12):5451--5460, 2019.

\bibitem{RLEStructuredMatrices}
J.~{Barbier}, N.~{Macris}, A.~{Maillard}, and F.~{Krzakala}.
\newblock {The Mutual Information in Random Linear Estimation Beyond i.i.d.
  Matrices}.
\newblock In {\em IEEE International Symposium on Information Theory (ISIT)},
  2018.

\bibitem{SM_Arxiv_Aubin2018}
B.~Aubin, A.~Maillard, J.~Barbier, F.~Krzakala, N.~Macris, and
  L.~Zdeborov{\'a}.
\newblock The committee machine: Computational to statistical gaps in learning
  a two-layers neural network.
\newblock In {\em Advances in Neural Information Processing Systems 31}, pages
  3227--3238, 2018.

\bibitem{Gabrie:NIPS2018}
M.~Gabri\'{e}, A.~Manoel, C.~Luneau, J.~Barbier, N.~Macris, F.~Krzakala, and
  L.~Zdeborov\'{a}.
\newblock Entropy and mutual information in models of deep neural networks.
\newblock In {\em Advances in Neural Information Processing Systems 31}, pages
  1824--1834. 2018.

\bibitem{coja2017information}
A.~Coja-Oghlan, F.~Krzakala, W.~Perkins, and L.~Zdeborov{\'a}.
\newblock Information-theoretic thresholds from the cavity method.
\newblock In {\em Proceedings of the 49th Annual ACM SIGACT Symposium on Theory
  of Computing (STOC)}, pages 146--157, 2017.

\bibitem{abbe2018community}
E.~Abbe.
\newblock Community detection and stochastic block models: Recent developments.
\newblock {\em Journal of Machine Learning Research}, 2018.

\bibitem{BarCM:2018}
J.~Barbier, C.~L. Chan, and N.~Macris.
\newblock Adaptive path interpolation for sparse systems: Application to a
  simple censored block model.
\newblock In {\em IEEE International Symposium on Information Theory (ISIT)},
  pages 1879--1883, 2018.

\bibitem{2019arXiv190106516B}
J.~Barbier and N.~Macris.
\newblock The adaptive interpolation method for proving replica formulas.
  applications to the curie{\textendash}weiss and wigner spike models.
\newblock {\em Journal of Physics A: Mathematical and Theoretical},
  52(29):294002, jun 2019.

\bibitem{macris2007griffith}
N.~Macris.
\newblock Griffith--kelly--sherman correlation inequalities: A useful tool in
  the theory of error correcting codes.
\newblock {\em IEEE Transactions on Information Theory}, 53(2):664--683, 2007.

\bibitem{MacrisKudekar2009}
S.~Kudekar and N.~Macris.
\newblock Sharp bounds for optimal decoding of low-density parity-check codes.
\newblock {\em IEEE Transactions on Information Theory}, 55(10):4635--4650, Oct
  2009.

\bibitem{andrea2008estimating}
A.~Montanari.
\newblock Estimating random variables from random sparse observations.
\newblock {\em Europ. Trans. on Telecomm.}, 19(4):385--403, 2008.

\bibitem{aizenman1998stability}
M.~Aizenman and P.~Contucci.
\newblock On the stability of the quenched state in mean-field spin-glass
  models.
\newblock {\em Journal of statistical physics}, 92(5-6):765--783, 1998.

\bibitem{contucci2005spin}
P.~Contucci and C.~Giardina.
\newblock Spin-glass stochastic stability: a rigorous proof.
\newblock In {\em Annales Henri Poincare}, volume~6. Springer, 2005.

\bibitem{panchenko2010ghirlanda}
D.~Panchenko.
\newblock The ghirlanda--guerra identities for mixed p-spin model.
\newblock {\em Comptes Rendus Mathematique}, 348(3-4):189--192, 2010.

\bibitem{schwarze1992generalization}
H.~Schwarze and J.~Hertz.
\newblock Generalization in a large committee machine.
\newblock {\em EPL (Europhysics Letters)}, 20(4):375, 1992.

\bibitem{monasson1995weight}
R.~Monasson and R.~Zecchina.
\newblock Weight space structure and internal representations: a direct
  approach to learning and generalization in multilayer neural networks.
\newblock {\em Physical review letters}, 75(12):2432, 1995.

\bibitem{engel2001statistical}
A.~Engel and C.~P. Van~den Broeck.
\newblock {\em Statistical Mechanics of Learning}.
\newblock Cambridge University Press, 2001.

\bibitem{panchenko2018potts}
D.~Panchenko.
\newblock Free energy in the potts spin glass.
\newblock {\em The Annals of Probability}, 46(2):829--864, 2018.

\bibitem{panchenko2018free}
D.~Panchenko.
\newblock Free energy in the mixed $ p $-spin models with vector spins.
\newblock {\em The Annals of Probability}, 46(2):865--896, 2018.

\bibitem{agliari2018non}
E.~Agliari, D.~Migliozzi, and D.~Tantari.
\newblock Non-convex multi-species hopfield models.
\newblock {\em Journal of Stat. Phys.}, 172(5):1247--1269, 2018.

\bibitem{manoel_multi-layer_2017}
A.~Manoel, F.~Krzakala, M.~Mézard, and L.~Zdeborov{\'a}.
\newblock {Multi-layer generalized linear estimation}.
\newblock In {\em IEEE International Symposium on Information Theory (ISIT)},
  2017.

\bibitem{reeves_additivity_2017}
G.~Reeves.
\newblock {Additivity of Information in Multilayer Networks via Additive
  Gaussian Noise Transforms}.
\newblock In {\em 55th Annual Allerton Conference on Communication, Control,
  and Computing}, 2017.

\bibitem{fletcher_inference_2017}
A.~K. Fletcher and S.~Rangan.
\newblock {Inference in Deep Networks in High Dimensions}.
\newblock {\em arXiv:1706.06549}, 2017.

\bibitem{DBLP:journals/corr/abs-1903-01293}
P.~Pandit, M.~Sahraee, S.~Rangan, and A.~K. Fletcher.
\newblock Asymptotics of map inference in deep networks.
\newblock {\em preprint arXiv:1903.01293}, 2019.

\bibitem{aubin2019spiked}
B.~Aubin, B.~Loureiro, A.~Maillard, F.~Krzakala, and L.~Zdeborov{\'a}.
\newblock The spiked matrix model with generative priors.
\newblock {\em preprint arXiv:1905.12385}, 2019.

\bibitem{horn1990matrix}
R.~A. Horn and C.~R. Johnson.
\newblock {\em Matrix analysis}.
\newblock Cambridge university press, 1990.

\bibitem{barbier2019overlap}
J.~Barbier.
\newblock Overlap matrix concentration in optimal bayesian inference.
\newblock {\em preprint arXiv:1904.02808}, 2019.

\end{thebibliography}

\end{document}